\documentclass[showpacs,preprintnumbers,amsmath,amssymb]{revtex4}

\usepackage{epsfig}
\newcommand{\be}{\begin{equation}}
\newcommand{\ee}{\end{equation}}
\newcommand{\bea}{\begin{eqnarray}}
\newcommand{\eea}{\end{eqnarray}}
\newcommand{\beaa}{\begin{eqnarray*}}
\newcommand{\eeaa}{\end{eqnarray*}}

\newcommand{\nn}{\nonumber \\}
\newcommand{\e}{{\rm e}}

\begin{document}

\tolerance=5000

\title{Unifying phantom inflation with late-time acceleration: \\
scalar phantom-non-phantom transition model \\
and generalized holographic dark energy}

\author{Shin'ichi Nojiri}\thanks{Electronic address: {\bf
 snojiri@yukawa.kyoto-u.ac.jp}}
\affiliation{Department of Physics,
Nagoya University, Nagoya 464-8602, Japan}
\author{Sergei D.~Odintsov}\thanks{Electronic address: {\bf
 odintsov@ieec.uab.es}, also at Lab.Fundamental Studies, TSPU, Tomsk}
\affiliation{Instituci\`o Catalana de Recerca i Estudis
Avan\c{c}ats (ICREA)  and Institut d'Estudis Espacials de Catalunya
(IEEC/ICE),
Edifici Nexus, Gran Capit\`a 2-4, 08034 Barcelona, Spain}

\date{\today}

\vskip 1pc
\begin{abstract}

The unifying approach to early-time and late-time universe based on
phantom cosmology is proposed. We consider gravity-scalar system which
contains usual potential and scalar coupling function in front of kinetic
term. As a result, the possibility of phantom-non-phantom  transition
appears in such a way that universe could have effectively phantom
equation of state at early time as well as at late time.
In fact, the oscillating universe may have several phantom and non-phantom phases.
The scalar factor in the kinetic term does not play any role
in each of two phase and can be absorbed into the redefinition
of the scalar field.  Right on the transition point, however, the factor cannot be
absorbed into the redefinition and play the role to connect two phases smoothly.
As a second model we suggest generalized holographic dark energy
where infrared cutoff is identified with combination of FRW parameters:
Hubble constant, particle and future horizons, cosmological constant and
universe life-time (if finite). Depending on the specific choice of the
model the number of interesting effects occur: the possibility to solve
the coincidence problem, crossing of phantom divide and unification of
early-time inflationary and late-time accelerating phantom universe.
The bound for holographic entropy which decreases in phantom era
is also discussed.

\end{abstract}


\maketitle
\vskip 1pc

\section{Introduction}

The recent astrophysical data indicate that effective equation of state
parameter $w_{\rm eff}$ of dark energy lies in the interval:
$-1.48 <w_{\rm eff}<-0.72$ \cite{HM}. In other words, it is quite possible
that current universe lives (or enters) at effective phantom era (for
review of observational data indicating to phantom late universe
see\cite{obse} and last ref. from \cite{HM} and for recent
discussion of various approaches to late-time phantom cosmology, see
\cite{phantom,tsujikawa} and references therein).
However, it is not clear how to relate the late-time phantom cosmology
with early-time inflation. For instance, the transition from
decelerating phase to dark energy universe is not yet well understood
(possibly because it is not clear what is dark energy itself).
Nevertheless, there are attempts to unify the early time inflation
where phantoms are essential with accelerated phantom universe
\cite{unification}. The unified inflation/acceleration universe
occurs for some models of modified gravity\cite{modified} as well
as for  complicated, non-standard equation of state (EOS) for the
universe \cite{salvatore} (for recent discussion of such (phantomic) EOS,
see \cite{EOS,inh}). The attempts to use phantoms in early universe may be
found also in \cite{inflation}.

In the present work we suggest the scenario where within the same theory,
quite naturally there occurs both phenomena: early-time phantom inflation
and late-time phantom acceleration.
The circles of phantom-dominated and non-phantom dominated epoch
in such universe suggest that probably the universe is (partially)
oscillating. In the next section we consider gravity-scalar theory  with
scalar-dependent coupling in front of kinetic term and scalar potential.
We should note that the scalar factor in the kinetic term does not play any role
each in the phontom or non-phantom phase and can be absorbed into the redefinition
of the scalar field.  Right on the transition point, however, the factor cannot be
absorbed into the redefinition and play the role to connect two phases smoothly.
In the number of explicit examples it is demonstrated how transitions
between phantom and non-phantom phases occur and that two phases are smoothly
connected with each other. As a result, there occurs the universe which
contains at least two phantom phases corresponding to early time inflation
and late time acceleration. The bridge between phantom phases correspond
to the standard non-phantom cosmology (radiation/matter dominated,
expanding or shrinking one). In oscillating universe there may emerge
multiple phantom/non-phantom transitions (eras).

Section three is devoted to the study of generalized holographic dark
energy where infrared cutoff depends on the combination of Hubble
 rate, the particle and future event horizons, life-time of the universe
 and even cosmological constant.
Here, the analogy with AdS/CFT correspondence may be pointed out:
there also IR or UV cutoffs represent some combination (depending on the
order of the expansion in large N). It is shown that in such model quite
naturally the crossing of phantom divide occurs. When including dark
matter, the natural solution of coincidence problem follows. Finally, it
is demonstrated that the unification of phantom inflation with phantom
dark energy universe is also possible. Some summary and outlook are given
in
the last section.

\section{Phantom inflation and late-time acceleration in scalar theory}

In the present section we will discuss usual gravity with scalar field.
The possibility to have unified phantom inflation with phantom late-time
acceleration is shown. This occurs via phantom-non-phantom transition.
Let us start from the following action:
\be
\label{k1}
S=\int d^4 x \sqrt{-g}\left\{\frac{1}{2\kappa^2}R - \frac{1}{2}\omega(\phi)\partial_\mu \phi
\partial^\mu \phi  - V(\phi)\right\}\ .
\ee
Here $\omega(\phi)$ and $V(\phi)$ are functions of the scalar field $\phi$.
Such scalar theory reminds about self-coupled dilaton \cite{naftulin} or
about sigma-model.
The spatially-flat FRW metric is
\be
\label{k2}
ds^2 = - dt^2 + a(t)^2 \sum_{i=1}^3 \left(dx^i\right)^2\ .
\ee
The scalar field $\phi$ only depends on the time coordinate $t$.
Then the FRW equations are given by
\be
\label{k3}
\frac{3}{\kappa^2}H^2 = \rho\ ,\quad - \frac{2}{\kappa^2}\dot H= p + \rho\ .
\ee
Here the energy density $\rho$ and the pressure $p$ are
\be
\label{k4}
\rho = \frac{1}{2}\omega(\phi){\dot \phi}^2 + V(\phi)\ ,\quad
p = \frac{1}{2}\omega(\phi){\dot \phi}^2 - V(\phi)\ .
\ee
By combining (\ref{k3}) and (\ref{k4}), one obtains
\be
\label{k5}
\omega(\phi) {\dot \phi}^2 = - \frac{2}{\kappa^2}\dot H\ ,\quad
V(\phi)=\frac{1}{\kappa^2}\left(3H^2 + \dot H\right)\ .
\ee
The interesting case is that $\omega(\phi)$ and $V(\phi)$ are defined in
terms of
single function
$f(\phi)$ as
\be
\label{k6}
\omega(\phi)=- \frac{2}{\kappa^2}f'(\phi)\ ,\quad
V(\phi)=\frac{1}{\kappa^2}\left(3f(\phi)^2 + f'(\phi)\right)\ .\ee
Hence, the following solution may be presented
\be
\label{k7}
\phi=t\ ,\quad H=f(t)\ .
\ee
One can check the solution (\ref{k7}) satisfies the scalar field equation:
\be
\label{k8}
0=\omega(\phi)\ddot \phi + \frac{1}{2}\omega'(\phi){\dot\phi}^2 + 3H\omega(\phi)\dot\phi
+ V'(\phi)\ .
\ee
Then {\it any} cosmology defined by $H=f(t)$ in (\ref{k7}) can be realized
by (\ref{k6}).

As clear from the first equation  (\ref{k5}), when $\dot H$ is positive,
which corresponds to the phantom phase, $\omega$ should be negative,
that is, the kinetic term of the scalar field has non-canonical sign.
On the other hand, when $\dot H$ is negative, corresponding to the non-phantom phase,
$\omega$ should be positive and the sign of the kinetic term of the scalar field is canonical.
If we restrict in one of phantom or non-phantom phase, the function
$\omega(\phi)$ can be
absorbed into the field redefinition given by
\be
\label{kk1}
\varphi = \int^\phi d\phi \sqrt{\omega(\phi)}\ ,
\ee
in non-phantom phase or
\be
\label{kk2}
\varphi = \int^\phi d\phi \sqrt{-\omega(\phi)}\ ,
\ee
in phantom phase.
Usually, at least locally, Eq.(\ref{kk1}) or Eq.(\ref{kk2}) can be solved with respect to $\phi$
as $\phi=\phi(\varphi)$. Then the action (\ref{k1}) can be rewritten as
\be
\label{k1B}
S=\int d^4 x \sqrt{-g}\left\{\frac{1}{2\kappa^2}R \mp \frac{1}{2}\partial_\mu \varphi
\partial^\mu \varphi  - \tilde V(\varphi)\right\}\ .
\ee
Here
\be
\label{ANYa}
\tilde V(\varphi)\equiv V\left(\phi(\varphi)\right)\ .
\ee
In the sign $\mp$ of (\ref{k1B}), the minus sign corresponds to the non-phantom phase and the plus one to
the phantom phase.
Then both of $\omega(\phi)$ and $V(\phi)$ in the action (\ref{k1})
do not correspond to physical degrees of freedom but
only one combination given by $\tilde V(\varphi)$ has real freedom in each of the phantom or non-phantom phase
and defines the real dynamics of the system.
The redefinition (\ref{kk1}) or (\ref{kk2}), however, has a discontinuity
between two phases. When explicitly keeping $\omega(\phi)$,
the two phases are smoothly connected with each other (kind of phase
transitions).
Hence, the function $\omega(\phi)$ gives only redundant degree of freedom and does not correspond to the
extra degree of freedom of the system ( in the phantom or non-phantom
phase). It plays the important role just in the point
of the transition
between the phantom phase and non-phantom phase.
By using the redundancy of $\omega(\phi)$, in any physcally equivalent model, one may choose,
just for example, $\omega(\phi)$ as
$\omega(\phi)=\omega_0\left(\phi - \phi_0\right)$ with constants $\omega_0$ and $\phi_0$. If we further
choose $\omega_0$ to be positive, the region given by $\phi>\phi_0$ corresponds to the non-phantom phase,
the region $\phi<\phi_0$ to the phantom phase, and the point $\phi=\phi_0$ to the point of the transition
between two phases.

Since the second FRW equation is given by
\be
\label{FRW2k}
p=-\frac{1}{\kappa^2}\left(2\dot H + 3H^2\right)\ ,
\ee
by combining the first FRW equation,  the effective equation of state
parameter
$w_{\rm eff}$ looks as
\be
\label{FRW3k}
w_{\rm eff}=\frac{p}{\rho}= -1 - \frac{2\dot H}{3H^2}\ .
\ee
After this discussion, one may consider some toy models to realize
above phantom-non-phantom transition to unify the phantom inflation
with late-time phantom acceleration.

As a first example, we consider the following model
\be
\label{k9}
f(\phi)=\frac{\alpha}{3}(T_0+\phi)^3 - \beta (T_0 + \phi) + \gamma\ ,\quad
\gamma \equiv - \frac{\alpha}{3}T_0^3 + \beta T_0\ ,
\ee
with the constants $\alpha$, $\beta$, and $T_0$, which give
\bea
\label{k10}
\omega(\phi)&=& - \frac{2}{\kappa^2}\left\{\alpha(T_0 + \phi)^2 - \beta\right\}\ ,\nn
V(\phi) &=& \frac{3}{\kappa^2}\Bigl\{\frac{\alpha^2}{3}(T_0 + \phi)^6
  - 2\alpha\beta (T_0 + \phi)^4 + \alpha\gamma (T_0 + \phi)^3 \nn
&& + \left(\alpha + 3\beta^2\right) (T_0 + \phi)^2 - 2\beta\gamma (T_0 + \phi)
+ 3\gamma^2 \Bigr\}\ .
\eea
Then the solution can be given by
\be
\label{hp25k}
H=\frac{\alpha}{3}(T_0+t)^3 - \beta (T_0 + t) + \gamma\ ,\quad \phi=t\ ,
\ee
or
\be
\label{hp26k}
a=a_0\e^{\frac{\alpha}{12}\left(T_0 + t\right)^4 - \frac{\beta}{2}\left(T_0 + t\right)^2
+ \gamma \left(T_0 + t\right)}\ .
\ee
As $H$ vanishes at $t=0$, $a$ has a minimum there.
Then the universe is shrinking when $t<0$ and expanding when $t>0$.
Since
\be
\label{hp27k}
\dot H=\alpha (T_0 + t)^2 - \beta\ ,
\ee
$\dot H$ vanishes at
\be
\label{hp28k}
t=t_\pm \equiv -T_0 \pm \sqrt{\frac{\beta}{\alpha}}>0\ .
\ee
Hence, $w_{\rm eff}$  (\ref{FRW3k}) is greater than $-1$ when $t_-<t<t_+$
(non-phantom phase)
and less than $-1$ when $0<t<t_-$ or $t>t_+$ (phantom phase).
There  occurs the phantom inflation when $0<t<t_-$ and late-time
acceleration when $t>t_+$.
We should also note there does not occur the Big Rip singularity in the solution (\ref{hp26k}) and
$w_{\rm eff}$ goes to $-1$ in the limit of $t\to \infty$.
Thus, the model (\ref{k10}) may provide a unification of the inflation
generated by phantom and
the late time phantom acceleration of the universe.

As a second example, we consider the model given by
\be
\label{k11}
f(\phi)=h_0 + h_1 \sin (\nu \phi)\ ,
\ee
with constants $h_0$, $h_1$, and $\nu$, which give
\bea
\label{k12}
\omega(\phi)&=& - \frac{2h_1\nu}{\kappa^2}\cos (\nu\phi)\ ,\nn
V(\phi)&=&\frac{3}{\kappa^2}\left(3h_0^2 + 6h_0h_1 \sin(\nu \phi) + h_1 \nu \cos (\nu\phi)
+ h_1^2\sin^2 (\nu\phi)\right)\ .
\eea
The Hubble rate $H$ is given by
\be
\label{k13}
H=h_0 + h_1 \sin (\nu t)\ ,
\ee
which is oscillating. When $h_0>h_1>0$, $H$ is always positive and the universe is
expanding.
Since
\be
\label{k14}
\dot H = h_1\nu \cos (\nu t)\ ,
\ee
when $h_1\nu>0$, $w_{\rm eff}$  (\ref{FRW3k}) is greater than $-1$ (non-phantom phase) when
\be
\label{k15}
\left(2n - \frac{1}{2}\right)\pi < \nu t < \left(2n + \frac{1}{2}\right)\pi \ ,
\ee
and less than $-1$ (phantom phase) when
\be
\label{k16}
\left(2n + \frac{1}{2}\right)\pi < \nu t < \left(2n + \frac{3}{2}\right)\pi \ .
\ee
In (\ref{k15}) and (\ref{k16}), $n$ is an integer.
Hence, in the model (\ref{k12}), there occur multiply oscillations
between phantom and non-phantom phases. It could be that our universe
currently corresponds to late-time acceleration phase in such
oscillatory regime.

The third example is given by scalar function :
\be
\label{kkk1}
f(\phi)=h_0\left(\frac{1}{\phi} + \frac{1}{t_s - \phi}\right)\ ,
\ee
with constants $h_0$ and $t_s$, which give
\bea
\label{kkk2}
\omega(\phi)&=& - \frac{2h_0 t_s \left(2\phi - t_s\right)}{\kappa^2 \phi^2
\left(t_s - \phi\right)^2}\ ,\nn
V(\phi)&=& \frac{h_0 t_s \left\{ 2\phi + \left(3h_0 - 1\right)t_s \right\}}{\kappa^2
\phi^2 \left(t_s - \phi\right)^2 }\ .
\eea
Then the Hubble rate and the scale factor $a$ are given by,
\be
\label{kkk3}
H=h_0\left(\frac{1}{t} + \frac{1}{t_s - t}\right)\ ,\quad
a=a_0\left(\frac{t}{t_s -t}\right)^{h_0}\ ,
\ee
This was obtained from the two scalars model in \cite{tsujikawa}.
As $a=0$ at $t=0$, the universe  starts at $t=0$.
Note that there is a Big Rip type singularity at $t=t_s$.
Since
\be
\label{kkk4}
\dot H = \frac{h_0 t_s \left(2t - t_s\right)}{t^2\left(t_s - t\right)^2} \ ,
\ee
when $0<t<t_s/2$, the universe is in non-phantom phase but when $t_s/2<t<t_s$, it is in
phantom phase. In fact $w_{\rm eff}$  (\ref{FRW3k}) is greater than $-1$ when
$0<t<t_s/2$ and less than $-1$ when $t_s/2<t<t_s$, Hence, again the
unified phantom inflation/acceleration universe may emerge.

As a fourth example, we now consider
\be
\label{k17}
f(\phi)=h_0^2 \left( \frac{1}{t_0^2 - \phi^2} + \frac{1}{\phi^2 + t_1^2}\right)\ .
\ee
Here $h_0$, $t_0$, and $t_1$ are positive constants.It is assumed
$t_0>t_1$.
Then $\omega(\phi)$ and $V(\phi)$ are
\bea
\label{k18}
\omega(\phi)&=& - \frac{8h_0^2 \left(t_0^2 + t_1^2\right)\phi
\left(\phi^2 - \frac{t_0^2 - t_1^2}{2}\right)}
{\kappa^2\left(t_0^2 - \phi^2\right)^2\left(\phi^2 - t_1^2\right)^2}\ ,\nn
V(\phi)&=&\frac{h_0^2 \left(t_0^2 + t_1^2\right)}
{\kappa^2\left(t_0^2 - \phi^2\right)^2\left(\phi^2 - t_1^2\right)^2}
\left\{ 3h_0^2 \left(t_0^2 + t_1^2\right)
+ 4\phi \left(\phi^2 - \frac{t_0^2 - t_1^2}{2}\right)\right\}\ .
\eea
The Hubble rate $H$ and the scale factor $a(t)$ follow as
\bea
\label{k19}
H&=&h_0^2 \left( \frac{1}{t_0^2 - t^2} + \frac{1}{t^2 + t_1^2}\right)\ ,\nn
a&=&a_0\left(\frac{t+ t_0}{t_0 - t}\right)^{\frac{h_0^2}{2t_0}}
\e^{-\frac{h_0^2}{t_1}{\rm Arctan}\frac{t_1}{t}}\ .
\eea
Since $a=0$ at $t=-t_0$, one may regard $t=-t_0$ corresponds to the
creation of the universe.
Since
\be
\label{k20}
\dot H = \frac{4h_0^2 \left(t_0^2 + t_1^2\right) t
\left(t^2 - \frac{t_0^2 - t_1^2}{2}\right)}
{\left(t_0^2 - t^2\right)^2\left(t^2 - t_1^2\right)^2}\ ,
\ee
we find $H$ has two minimum at $t=t_\pm \equiv \pm \sqrt{\frac{t_0^2 - t_1^2}{2}}$ and at $t=0$,
$H$ has a local maximum.
Hence, phantom phase occurs when $t_-<t<0$ and $t>t_+$ and non-phantom
phase
when $t_0>t>t_-$ and $0<t<t_+$.
We also note that there is a Big Rip type singularity at $t=t_0$.

As is discussed in \cite{CNO}, the solution (\ref{k7}) is stable in the
phantom phase
but unstable in the non-phantom phase.
The instablity becomes very large when crossing $w=-1$.
In order to avoid this problem, one may consider two scalar model.
In case of one scalar model, the large instability occurs since
 the coefficient of the kinetic term
$\omega(\phi)$ in (\ref{k1}) vanishes at the crossing $w=-1$ point.
 In the two scalar model,
we can choose the corresponding coefficients do not vanish anywhere.
Then we may expect that such a divergence of the instability would
not occur, which we now check explicitly.

We now consider two scalar model like
\be
\label{A1}
S=\int d^4 x \sqrt{-g}\left\{\frac{1}{2\kappa^2}R - \frac{1}{2}\omega(\phi)\partial_\mu \phi
\partial^\mu \phi - \frac{1}{2}\eta(\chi)\partial_\mu \chi
\partial^\mu \chi - V(\phi,\chi)
\right\}\ .
\ee
Here $\eta(\chi)$ is a function of the scalar field $\chi$.
The FRW equations give
\be
\label{A2}
\omega(\phi) {\dot \phi}^2 + \eta(\chi) {\dot \chi}^2 = - \frac{2}{\kappa^2}\dot H\ ,\quad
V(\phi,\chi)=\frac{1}{\kappa^2}\left(3H^2 + \dot H\right)\ .
\ee
Then if
\be
\label{A3}
\omega(t) + \eta(t)=- \frac{2}{\kappa^2}f'(t)\ ,\quad
V(t,t)=\frac{1}{\kappa^2}\left(3f(t)^2 + f'(t)\right)\ ,
\ee
the explicit solution follows
\be
\label{A4}
\phi=\chi=t\ ,\quad H=f(t)\ .
\ee
One may choose that $\omega$ should be always positive and $\eta$ be
always negative, for example
\bea
\label{A5}
\omega(\phi)&=&-\frac{2}{\kappa^2}\left\{f'(\phi) - \sqrt{\alpha^2 + f'(\phi)^2}\right\}>0\ ,\nn
\eta(\chi)&=&-\frac{2}{\kappa^2}\sqrt{\alpha^2 + f'(\chi)^2}<0\ .
\eea
Here $\alpha$ is a constant. Define a new function $\tilde f(\phi,\chi)$
by
\be
\label{A6}
\tilde f(\phi,\chi)\equiv - \frac{\kappa^2}{2}\left(\int d\phi \omega(\phi)
+ \int d\chi \eta(\chi)\right)\ ,
\ee
which gives
\be
\label{A7}
\tilde f(t,t)=f(t)\ .
\ee
If $V(\phi,\chi)$ is given by using $\tilde f(\phi,\chi)$ as
\be
\label{A8}
V(\phi,\chi)=\frac{1}{\kappa^2}\left(3{\tilde f(\phi,\chi)}^2
+ \frac{\partial \tilde f(\phi,\chi)}{\partial \phi}
+ \frac{\partial \tilde f(\phi,\chi)}{\partial \chi} \right)\ ,
\ee
 the FRW  and the scalar field equations are also satisfied:
\bea
\label{A9}
0&=&\omega(\phi)\ddot\phi + \frac{1}{2}\omega'(\phi) {\dot \phi}^2
+ 3H\omega(\phi)\dot\phi + \frac{\partial \tilde V(\phi,\chi)}{\partial \phi}\ ,\nn
0&=&\eta(\chi)\ddot\chi + \frac{1}{2}\eta'(\chi) {\dot \chi}^2
+ 3H\eta(\chi)\dot\chi + \frac{\partial \tilde V(\phi,\chi)}{\partial \chi}\ .
\eea

In case of one scalar model, the instability becomes infinite
at the crossing $w=-1$ point (from higher than phantom value),
since the coefficient of the kinetic term $\omega(\phi)$ in (\ref{k1})
vanishes at the point. In the two scalar model in (\ref{A1}),
the coefficients $\omega(\phi)$
and $\eta(\phi)$ do not vanish anywhere, as in (\ref{A5}). Then we may expect that such
a divergence of the instability does not occur.

By introducing the new quantities, $X_\phi$, $X_\chi$, and $Y$ as
\be
\label{A10}
X_\phi \equiv \dot \phi\ ,\quad X_\chi \equiv \dot \chi\ ,\quad
Y\equiv \frac{\tilde f(\phi,\chi)}{H} \ ,
\ee
 the FRW equations and the scalar field equations (\ref{A9}) are:
\bea
\label{A11}
\frac{dX_\phi}{dN}&=& - \frac{\omega'(\phi)}{2H \omega(\phi)}\left(X_\phi^2 - 1\right)
 - 3(X_\phi-Y)\ ,\nn
\frac{dX_\chi}{dN}&=& - \frac{\eta'(\chi)}{2H \eta(\chi)}\left(X_\chi^2 - 1\right)
 - 3(X_\chi-Y)\ ,\nn
\frac{dY}{dN}&=&\frac{1}{2\kappa^2H^2}\left\{X_\phi \left(X_\phi Y -1\right)
+ X_\chi\left(X_\chi Y -1\right)\right\}\ .
\eea
Here $d/dN\equiv H^{-1}d/dt$. In the solution (\ref{A4}),  $X_\phi=X_\chi=Y=1$.
The following perturbation may be considered
\be
\label{A12}
X_\phi=1+\delta X_\phi\ ,\quad X_\chi=1 + \delta X_\chi\ ,\quad Y=1 + \delta Y\ .
\ee
Hence
\be
\label{A13}
\frac{d}{dN}\left(\begin{array}{c}
\delta X_\phi \\
\delta X_\chi \\
\delta Y
\end{array}\right)
= M \left(\begin{array}{c}
\delta X_\phi \\
\delta X_\chi \\
\delta Y
\end{array}\right)
\ ,\quad
M\equiv \left(\begin{array}{ccc}
- \frac{\omega'(\phi)}{H\omega(\phi)} - 3 & 0 & 3 \\
0 & - \frac{\eta'(\chi)}{H\eta(\chi)} - 3 & 3 \\
\frac{1}{2\kappa^2H^2} & \frac{1}{2 \kappa^2 H^2} & \frac{1}{\kappa^2 H^2}
\end{array}\right)\ .
\ee
The eigenvalues of the matrix $M$ are given by solving the following eigenvalue
equation
\bea
\label{A14}
0&=& \left(\lambda + \frac{\omega'(\phi)}{H\omega(\phi)} + 3\right)
\left(\lambda + \frac{\eta'(\chi)}{H\eta(\chi)} + 3\right)
\left(\lambda - \frac{1}{\kappa^2 H^2}\right) \nn
&& + \frac{3}{2\kappa^2 H^2}\left(\lambda + \frac{\omega'(\phi)}{H\omega(\phi)} + 3\right)
+ \frac{3}{2\kappa^2 H^2}\left(\lambda + \frac{\eta'(\chi)}{H\eta(\chi)} + 3\right)\ .
\eea
The eigenvalues (\ref{A14}) for the two scalar model are clearly finite.
Hence, the instability could be finite.
In fact, right on the transition point where $\dot H=f'(t)=0$ and therefore $f'(\phi)=f'(\chi)=0$,
for the choice in (\ref{A5}), we find
\be
\label{AA1}
\omega(\phi)=-\eta(\chi)=\frac{2\alpha}{\kappa^2}\ ,\quad
\omega'(\phi)=-\frac{2\ddot H}{\kappa^2}\ ,\quad \eta'(\chi)=0\ .
\ee
Then the eignvalue equation (\ref{A14}) reduces to
\be
\label{AA2}
0=\lambda^3 + \left(-A-B + 6\right) \lambda^2 + \left(AB - 3A - 3B + 9\right) \lambda
 - \frac{3}{2}AB + 9B\ ,\quad
A\equiv \frac{\ddot H}{\alpha}\ ,\quad B\equiv \frac{1}{\kappa^2 H^2}\ .
\ee
Here we have chosen $\alpha>0$. Then the eignevalues are surely finite, which tells that even if
the solution (\ref{A4}) could not be stable, the solution has non-vanishing measure and therefore
the transition from non-phantom phase to phantom one can surely occur.
We should also note that the solution (\ref{A4}) can be in fact stable. For example, we consider the case
$A,B\to 0$. Then Eq.(\ref{AA2}) further reduces to
\be
\label{AA3}
0=\lambda\left(\lambda + 3\right)^2\ .
\ee
Then the eignvalues are given by $0$ and $-3$. Since there is no positive eigenvalue, the solution
(\ref{A4}) is stable in the case.

As an example, we consider $f(t)=h_0 + h_1 \sin(\nu t)$ in (\ref{k11}).
 Here it is assumed
$h_0$, $h_1$, and $\nu$ are positive. By choosing $\alpha = h_1\nu$
in (\ref{A5}), one finds
\bea
\label{Ae1}
&& \omega(\phi)=-\frac{2h_1\nu}{\kappa^2}\left\{\cos(\nu \phi) - \sqrt{1 + \cos^2(\nu \phi)}\right\}\ ,\quad
\eta(\chi)=-\frac{2h_1}\nu{\kappa^2}\sqrt{1 + \cos^2(\nu \chi)}\ ,\nn
&& \tilde f(\phi,\chi) = h_0 + h_1 \sin(\nu\phi) - \frac{\sqrt{2}}{\nu}\left\{
E\left(1/\sqrt{2}, \nu \phi\right) - E\left(1/\sqrt{2}, \nu \chi\right)\right\}\ .
\eea
Here $E(k,x)$ is the second kind elliptic integral defined by
\be
\label{Ae2}
E(k,x)=\int_0^x dx \sqrt{1 - k^2 \sin^2 x}\ .
\ee
Then even in two scalar model, the cosmology is given by (\ref{k13}).
Similarly  any cosmology (including unified inflation/acceleration) can be
realized by using the two scalar model, as
in the examples with single scalar field.

Thus, we presented several toy models showing the natural possibility
to unify early-time inflation with late-time acceleration via
phantom-non-phantom transitions in scalar theory. Much work remains to be
done in order to decide if such theoretic possibility is realistic one.
In next section, we demonstrate that similar effect is possible also in
generalized holographic dark energy.

\section{Generalized holographic dark energy  and unification
of phantom inflation with phantom acceleration.}

Let us start from  the holographic dark energy model \cite{Li} (see also
refs.\cite{FEH} where further support for holographic DE was given).
Denote the
infrared cutoff by
$L_\Lambda$, which has a dimension of length. If the holographic
dark energy $\rho_\Lambda$ is given by,
\be
\label{H1}
\rho_\Lambda=\frac{3c^2}{\kappa^2 L_\Lambda^2}\ ,
\ee
with a numerical constant $c$, the first FRW equation
\be
\label{FRW1}
\frac{3}{\kappa^2}H^2=\rho_\Lambda\ ,
\ee
can be written as
\be
\label{H2}
H=\frac{c}{L_\Lambda}\ .
\ee
Here it is assumed that $c$ is positive
to assure the expansion of the universe.
In (\ref{FRW1}), we do not include the contribution from the matter.
The next question is the choice of infrared cut-off.
For instance, identifying it with Hubble parameter does not lead
to accelerating universe.
Hence, one is forced to consider other choices.

The particle horizon $L_p$
and future horizon $L_f$ are defined by
\be
\label{H3}
L_p\equiv a
\int_0^t\frac{dt}{a}\ ,\quad L_f\equiv a \int_t^\infty \frac{dt}{a}\ .
\ee
For the FRW metric with the flat spacial part:
\be
\label{H4}
ds^2 = -dt^2 + a(t)^2\sum_{i=1,2,3}\left(dx^i\right)^2\ .
\ee
Identifying $L_\Lambda$ with $L_p$ or $L_f$, one obtains the
following equation:
\be
\label{H5}
\frac{d}{dt}\left(\frac{c}{aH}\right)=\pm \frac{1}{a}\ .
\ee
Here, the plus (resp. minus) sign corresponds to the particle (resp.
future) horizon. The solution of (\ref{H5}) is given by
\be
\label{H6}
a=a_0 t^{h_0}\ ,
\ee
with
\be
\label{H7}
h_0=\frac{1}{1\pm \frac{1}{c}}\ .
\ee
Then, in the case $L_\Lambda=L_f$, the universe is accelerating ($h_0>1$ or $w=-1 +
{2}/{3h_0}<-1/3$). When $c>1$ in the case $L_\Lambda=L_p$, $h_0$
becomes negative and the universe is shrinking. If the theory is
invariant under the change of the direction of time, one may change
$t$ with $-t$. Furthermore by properly shifting the origin of time,
we obtain, instead of (\ref{H6}),
\be
\label{H8}
a=a_0\left(t_s - t\right)^{h_0}\ .
\ee
This tells us that there will be a Big Rip
singularity at $t=t_s$ (for classification of future, finite-time
singularities and list of related references, see \cite{tsujikawa}).
  Since
the direction
of time is changed, the particle horizon becomes a future like one:
\be
\label{H8b}
L_p\to \tilde L_f \equiv a \int_t^{t_s}\frac{dt}{a}=a \int_0^\infty
\frac{da}{Ha^2}\ .
\ee
By using (\ref{FRW3k}) for (\ref{H6}) and (\ref{H8}), we find
\be
\label{FRW4}
w_{\rm eff}= -1 + \frac{2}{3h_0}\ .
\ee

Note that if  $L_\Lambda$ is chosen as a future horizon in
(\ref{H3}), the deSitter space
\be
\label{dS1}
a=a_0\e^{\frac{t}{l}}\quad \left(H=\frac{1}{l}\right)
\ee
can be a solution. Since $L_f$ is now given by $L_f=l$, the
holographic dark energy  (\ref{H1}) is given by
$\rho_\Lambda=\frac{c^2}{\kappa^2 l^2}$. When $c=1$, the first
FRW equation $\frac{3}{\kappa^2} H^2 = \rho_\Lambda$ is identically
satisfied. If $c\neq 1$, the deSitter space is not a solution. If
  $L_\Lambda$ is chosen to be the particle horizon, the deSitter solution
does not exist, either, since the particle horizon $L_p$  (\ref{H3}) is
not a constant: $L_p=(1 - \e^{\frac{t}{l}} )/l$.
Hence, the essentials of holographic dark energy are discussed.

In general, $L_\Lambda$ could be a combination (a function) of both,
$L_p$, $L_f$\cite{eli}. Furthermore, if the span of life of the universe
is
finite, the span $t_s$ can be an infrared cutoff. If the span of
life of the universe is finite, the definition of the future horizon
$L_f$  (\ref{H3}) is not well-posed, since $t$ cannot go to
infinity. Then, one may redefine the future horizon as in (\ref{H8b})
\be
\label{H8c}
L_f\to \tilde L_f \equiv a \int_t^{t_s}\frac{dt}{a}=a \int_0^\infty \frac{da}{Ha^2}\ .
\ee
Since there can be many choices for the infrared cutoff,
in analogy with AdS/CFT one may
assume $L_\Lambda$ is the function of $L_p$, $\tilde L_f$, and
$t_s$, as long as these quantities are finite:
\be
\label{H9}
L_\Lambda=L_\Lambda\left(L_p, \tilde L_f, t_s\right)\ .
\ee
As an example, we consider the generalized holographic dark energy
from above class \cite{eli}
\be
\label{H10}
\frac{L_\Lambda}{c}=\frac{ \left( \frac{t_s B\left( 1+h_0,1-h_0 \right)}{L_p + \tilde L_f}\right)^{1/h_0}}{ h_0
\left\{ 1 + \left(\frac{t_s B\left(1+h_0,1-h_0\right)}{L_p + \tilde L_f}\right)^{1/h_0}\right\}^2}
\ ,\quad h_0>0\ .
\ee
Here $B(p,q)$ is a beta-function defined by
\be
\label{hp1}
B(p,q)\equiv \int_0^\infty \frac{dt\,t^{p-1}}{\left(1+t\right)^{p+q}}\ .
\ee
  Eq.(\ref{H10}) leads to the solution:
\be
\label{H11}
H=h_0\left(\frac{1}{t} + \frac{1}{t_s - t}\right)\ ,\quad \mbox{or}
\quad a=a_0 \left(\frac{t}{t_s - t}\right)^{h_0}\ .
\ee
In fact, one finds
\be
\label{H12}
L_p+\tilde L_f = a \int_0^{t_s} \frac{dt}{a}= t_s
\left(\frac{t}{t_s - t}\right)^{h_0} B\left(1+h_0, 1-h_0\right)\ ,
\ee
and therefore
\be
\label{H13}
\frac{c}{L_\Lambda}=h_0\left(\frac{1}{t} + \frac{1}{t_s -
t}\right)=H\ ,
\ee
which satisfies (\ref{H2}).
For the solution (\ref{H11}),  $w_{\rm eff}$ defined in (\ref{FRW3k})
  is time-dependent and looks as
\be
\label{FRW5}
w_{\rm eff}= - 1 + \frac{2\left( t_s - 2t\right)}{3h_0 t_s}\ .
\ee
Then $w_{\rm eff}=-1$ at $t=t_s/2$ and we find $w\to -1 + 2/(3h_0)>-1$ when $t\to 0$ and
$w_{\rm eff}\to -1 - 2/(3h_0)<-1$ when $t\to t_s$. Hence, there occurs the
crossing of
$w_{\rm eff}=-1$ in our generalized holographic dark energy model.

One can also include the matter whose parameter of the equation of state
is $w$: $\rho_m = wp_m$ into the consideration. In the following, we define $h_0$ 
by $h_0\equiv (2/3)/(1+w)$. 
However, it is assumed that
  there is an interaction between the holographic matters\cite{amendola}.
The energy-conservation law is taken as follows
\be
\label{hp2}
\dot\rho_m + 3H\left(\rho_m + p_m\right)=3H \cdot
\frac{4\rho_0}{3h_0}\frac{\left\{
1 + \left(\frac{t_s B\left(1+h_0,1-h_0\right)}{L_p + \tilde L_f}\right)^{1/h_0}\right\}^3}
{\left(\frac{t_s B\left(1+h_0,1-h_0\right)}{L_p + \tilde L_f}\right)^{2/h_0}}\ .
\ee
It is also assumed that
\be
\label{hp3}
\frac{L_\Lambda}{c} =\left( 1 - \frac{\kappa^2 \rho_0}{3h_0^2}\right)^{-1/2}
\frac{\left(\frac{t_s B\left(1+h_0,1-h_0\right)}{L_p + \tilde L_f}\right)^{1/h_0}}{h_0\left\{
1 + \left(\frac{t_s B\left(1+h_0,1-h_0\right)}{L_p + \tilde L_f}\right)^{1/h_0}\right\}^2}\ ,
\ee
The first FRW equation is then modified as
$(H^2/\kappa^2)=\rho_\Lambda + \rho_m$ and we find (\ref{H11}) again and
\be
\label{hp4}
\rho_m = \rho_0 \left(\frac{1}{t} + \frac{1}{t_s -t}\right)^2\ .
\ee
Then the ratio of the energy density $\rho_m$ of matter with respect to that of
the holographic dark energy corresponding to (\ref{H1}) is a constant:
\be
\label{hp5}
\frac{\rho_\Lambda}{\rho_m}=\frac{3h_0^2}{\kappa^2\rho_0}
\left(1 - \frac{\kappa^2 \rho_0}{3h_0^2}\right)\ .
\ee
Hence, one sees that coincidence problem may be solved in
generalized holographic dark energy. Note that similar scenario
was proposed in \cite{PZ}, where the ratio of the matter energy density
and the holographic energy
can be a constant by introducing the interaction, as in (\ref{hp2}), between the matter and the
holographic energy. For the naive model  \cite{PZ},
  the effective $w_{\rm eff}$
  (\ref{FRW3k}) is constant. As shown here, even if $w_{\rm eff}$ depends on time, the ratio
can be constant.
In more general case the ratio between the matter energy density and the
dark energy density
is not constant  \cite{WGA}.

As more general case than (\ref{H9}), we may consider the case that $L_\Lambda$ depends on
the Hubble rate $H$ and the length scale $l$ coming from the cosmological constant
$\Lambda=12/l^2$, if the cosmological constant does not vanish:
\be
\label{hp6}
L_\Lambda=L_\Lambda\left(L_p, \tilde L_f, t_s, H, l\right)
\ \mbox{or}\ L_\Lambda\left(L_p, L_f, t_s, H, l\right)\ .
\ee
As an example, the generalized holographic dark energy theory from such
class
may be considered
\be
\label{hp7}
\frac{c}{L_\Lambda}=\frac{1}{\alpha L_f}\left\{ \alpha + 1 +
2\left(\alpha^2 - \alpha -1\right)\frac{L_f}{\alpha l}
+ 2\left(\alpha^3 - 2\alpha^2 + \alpha + 1\right)\left(\frac{L_f}{\alpha l}\right)^2 \right\}\ .
\ee
Here $\alpha$ is a positive dimensionless parameter.
Since
\bea
\label{hp7b}
&& \left(\alpha^2 - \alpha -1\right)^2 - 2\left(\alpha^3 - 2\alpha^2 + \alpha + 1\right)
= - \left(\alpha^2 - \frac{1}{2}\right)^2 - 2\alpha - \frac{3}{4}<0 \nn
&& \alpha^3 - 2\alpha^2 + \alpha + 1=\alpha (\alpha -1)^2 + 1 > 0\ ,
\eea
$L_\Lambda$ is always positive as long as $\alpha$ is positive.
Then a cosmological solution is given by
\be
\label{hp8}
a=\frac{t^{\alpha +1}\e^{\frac{t}{l}}}{L_0\alpha \left( 1 + \frac{t}{\alpha l}\right)}\ ,
\quad L_f=\frac{t}{\alpha\left(1 + \frac{t}{\alpha l}\right)}\ ,
\ee
which leads to
\be
\label{hp9}
H=\frac{1 + \alpha \left(1 + \frac{t}{\alpha l}\right)^2}{t\left(1 + \frac{t}{\alpha l}\right)} .
\ee
As clear from (\ref{hp8}), if $\alpha<0$, there occurs the Big Rip like singularity at
$t=-\alpha l$, where $a$ diverges. When $\alpha<0$, $L_\Lambda$ can be
negative, in such a case as
$H=\frac{c}{L_\Lambda}$ if we do not include the matter, the universe is
shrinking.

In (\ref{hp8}), $L_0$ is a constant of the integration.
When $t$ is small, $H$ behaves as the inverse power of $t$:
\be
\label{hp10}
H\to \frac{\alpha + 1}{t}\ ,
\ee
which shows that the universe is filled with the fluid with
$w=-(\alpha - 1)/(3(\alpha + 1))$.
On the other hand, when $t$ is large, $H$ goes to a constant:
\be
\label{hp11}
H\to \frac {1}{l}\ ,
\ee
which tells that the universe becomes deSitter space asymptotically.
Instead of (\ref{hp7}), one may consider the model as
\be
\label{hp12}
\frac{c}{L_\Lambda}=\frac{H}{\alpha + 1} + \frac{1}{L_f}\left\{ 1
+ \frac{2\left(\alpha^2 - \alpha - 1\right)}{\alpha + 1}\frac{L_f}{\alpha l}
+ \left(1 - \frac{\alpha^2(\alpha + 2)}{\alpha + 1} \right)
\left(\frac{L_f}{\alpha l}\right)^2 \right\}\ .
\ee
Then the solution (\ref{hp8}) follows again.

In general, the FRW equation has the following form:
\be
\label{hp13}
\frac{3}{\kappa^2}H^2 = \frac{3c^2}{\kappa^2 L_\Lambda^2} + \rho_m\ .
\ee
If one defines the matter energy $E_m$, the Casimir energy $E_c$, and the
Hubble entropy
$S_H$ by
\be
\label{hp14}
E_m\equiv \rho_m L_\Lambda^3\ ,\quad
E_c\equiv \frac{3c^2L_\Lambda}{\kappa^2}\ ,\quad
S_H\equiv \frac{18\pi cL_\Lambda^3}{\kappa^2}H\ ,
\ee
we can rewrite the FRW equation in a Cardy-Verlinde  (holographic) form
\cite{verlinde}:
\be
\label{hp15}
S_H^2=\left( 2\pi L_\Lambda\right)^2E_c\left(E_c + E_m\right)\ .
\ee
Different from the deSitter case \cite{KeLi}, the Hubble entropy $S_H$ is
not a constant and
depends on time.
In case $\rho_m=0$, by using (\ref{H2}) one has
\be
\label{hp16}
S_H= \frac{18\pi c^2 L_\Lambda^2}{\kappa^2}= \frac{18\pi c^4}{\kappa^2 H^2}\ .
\ee
Thus for the model (\ref{H10}),  using (\ref{H11}), we find
\be
\label{hp17}
S_H=\frac{18\pi c^4t^2 \left(t_s - t\right)^2}{\kappa^2 h_0^2 t_s^2}\ ,
\ee
which vanishes at $t=0$ and $t=t_s$. The maximum of $S_H$ (\ref{hp17}) is obtained when $t=t_s/2$.
For the generalized models (\ref{hp2}) and (\ref{hp3}), where matter is
included, $S_H$  (\ref{hp17}) is modified
by a constant factor:
\be
\label{hp18}
S_H=\left( 1 - \frac{\kappa^2 \rho_0}{3h_0^2}\right)^{-3}
\frac{18\pi c^4t^2 \left(t_s - t\right)^2}{\kappa^2 h_0^2 t_s^2}\ .
\ee
In case of (\ref{hp7}) or (\ref{hp13}), it follows
\be
\label{hp19}
S_H=\frac{18\pi c^4 t^2 \left(1 + \frac{t}{\alpha l}\right)^2}
{\kappa^2\left\{ 1 + \alpha \left(1 + \frac{t}{\alpha l}\right)^2\right\}^2}\ ,
\ee
which vanishes at $t=0$ and goes to a constant
\be
\label{hp20}
S_H\to \frac{18\pi c^4 l^2}{\kappa^2}\ ,
\ee
when $t\to \infty$.
In (\ref{hp19}), $\alpha$ can be negative in general but $S_H$ is positive.
Although the Hubble entropy $S_H$ is always positive in (\ref{hp16}),
  $S_H$  (\ref{hp18}) can be negative
if $ \frac{\kappa^2 \rho_0}{3h_0^2}>1$. In such a case, if $S_H$ gives a upper bound for the entropy of
the universe, the entropy should be negative. Such a negative entropy
has been observed for phantom era in \cite{BNOV}. Nevertheless,
when phantom era is transient in the late-time universe it may occur
that the universe entropy remains to be positive \cite{inh}.

As a little bit complicated example, we may consider
\be
\label{hp21}
\frac{c}{L_\Lambda}=\frac{\alpha}{3}T^3 - \beta T + \gamma\ .
\ee
Here $\alpha$, $\beta$, and $\gamma$ are positive constants satisfying
\be
\label{hp22}
\gamma>\frac{2\beta}{3}\sqrt{\frac{\beta}{\alpha}}\ ,
\ee
and $T$ is defined by
\bea
\label{hp23}
T&\equiv &\frac{H-\gamma + \frac{4\alpha}{3\beta^2}\left(H + 3\gamma\right)\ln \frac{L}{F}}
{\frac{\alpha}{3\beta^2}\left(H + 3\gamma\right)^2 - \beta
  - \frac{4\alpha}{3\beta}\ln\frac{L}{F}}\ ,\nn
L&\equiv & \frac{a(t)}{a(0)} F\ ,\nn
F&\equiv & \int_{-\infty}^\infty \e^{-\frac{\alpha}{12}t^4 + \frac{\beta}{2}t^2 - \gamma t} dt\ .
\eea
The condition (\ref{hp23}) tells that, as a function of $T$, $H$ vanishes only once.
By defining $T_0<0$ via $H(T=T_0)=0$, one gets
\be
\label{hp24}
\gamma = - \frac{\alpha}{3}T_0^3 + \beta T_0\ .
\ee
Then the solution can be given by (\ref{hp25k}) or (\ref{hp26k}), again.
Thus, the generalized dark energy model (\ref{hp21}) may provide a
unification of the inflation generated by phantom and
the late time phantom acceleration of the universe.

In the model (\ref{hp21}), the Hubble entropy $S_H$
  (\ref{hp14}) is given by
\be
\label{hp29}
S_H=\frac{18\pi c^4}{\kappa^2\left\{\frac{\alpha}{3}\left(t+T_0\right)^3
  - \beta \left(t+T_0\right) + \gamma\right\}^2}\ ,
\ee
which is always positive. From (\ref{hp24}), one finds $S_H$ diverges at
$t=0$. Since $S_H$ is proportional to $H^{-2}$,
$S_H$ decreases when $0<t<t_-$, increases when $t_-<t<t_+$, and decreases
again when $t>t_+$.
In the limit $t\to \infty$, $S_H$ vanishes.
Therefore, in the phantom phase, $S_H$ is decreasing and in the
non-phantom phase, it is increasing. This finishes the discussion
of holographic entropy bounds for generalized holographic dark energy
model.

\section{Discussion}

In summary, it is suggested the scenario where phantom cosmology may be key
element at early-time as well as at late-time universe.
Specific model under consideration suggests the phantom-non-phantom
transitions during the evolution of the universe. This may be easily
realized due to presence of scalar coupling in front of kinetic term:
this coupling function may change its sign on cosmological scales.
As a result, even multiply phantom-non-phantom transitions are possible.
The intermediate region between very early and very late universe may
correspond to standard (radiation/matter dominated) cosmology.

The generalized holographic dark energy where infrared cutoff is
identified with combination of the natural FRW parameters:
Hubble rate, particle and future horizons, span of life of the universe
(when its life-time is finite) and even with cosmological constant is
suggested.
The possibility to have the crossing of phantom divide there, as well as
solution of coincidence problem (when matter presents) is demonstrated.
The holographic entropy bound is obtained and the regime where it may be
negative is discussed. It is interesting that holographic entropy is
decreasing in phantom phase in accord with earlier observation of
ref.\cite{BNOV}. Finally, the possibility to unify phantom inflation with
late-time acceleration is demonstrated also in generalized holographic
dark energy. It is clear that in similar way one can suggest unified
cosmological scenario for tachyon phantoms and for
time-dependent, phantomic equations of
state.

Much work remains to be done in order to understand if such combined
scenario is realistic one: the standard cosmological problems (especially
at early universe) should be first discussed. In order to start such
investigation one awaits the final answer to fundamental question:
Do we currently live in phantom universe? Hopefully, the answer comes
soon.

\section*{Acknowledgments}

The research by SDO has been partly supported by RFBR grant 06-01-00609
(Russia),
by LRSS project n4489.2006.02 (Russia) and by project FIS2005-01181
(MEC, Spain).


\begin{thebibliography}{99}


\bibitem{HM}
S.~Hannestad and E.~Mortsell,
Phys.\ Rev.\ D {\bf 66}, 063508 (2002);

A.~Melchiori, L.~Mersini-Houghton, C.~J.~Odman, and M.~Trodden,
Phys.\ Rev.\ D {\bf 68}, 043509 (2003);

H.~Jassal, J.~Bagla and T.~Padmanabhan,
arXiv:astro-ph/0506748.
\bibitem{obse}
L.Perivolaropoulos, astro-ph/0601014.

\bibitem{phantom}
R.~R.~Caldwell, M.~Kamionkowski, and N.~N.~Weinberg,
Phys.\ Rev.\ Lett.\  {\bf 91}, 071301 (2003)
[arXiv:astro-ph/0302506];

B.~McInnes,
JHEP {\bf 0208}, 029 (2002)
[arXiv:hep-th/0112066];

V.~Faraoni,
Int.\ J.\ Mod.\ Phys.\ D {\bf 64}, 043514 (2002);
gr-qc/0404078; gr-qc/0506095;

S.~Nojiri and S.~D.~Odintsov,
Phys.\ Lett.\ B {\bf 565}, 1 (2003)
[arXiv:hep-th/0304131];
Phys.\ Lett.\ B {\bf 595}, 1 (2004)
[arXiv:hep-th0405078];

P.~Singh, M.~Sami, and N.~Dadhich,
Phys.\ Rev.\ D {\bf 68}, 023522 (2003)
[arXIv:hep-th/0305110];

P.~Gonzalez-Diaz,
Phys.\ Lett.\ {\bf B586}, 1 (2004)
[arXiv:astro-ph/0312579];
arXiv:hep-th/0408225;

H.~Stefancic,
Eur.\ Phys.\ J.\ C {\bf 36}, 523 (2004)
[arXiv:astro-ph/0312484];

M.~Sami and A.~Toporensky,
Mod.\ Phys.\ Lett.\ {\bf A19}, 1509 (2004)
[arXiv:gr-qc/0312009];

X.~Meng and P.~Wang,
Class.\ Quant.\ Grav.\ {\bf}, 22 (2005)
[arXiv:gr-qc/0411007];



S.~M.~Carroll, A.~De~Felice, and M.~Trodden,
Phys.\ Rev.\ D {\bf 71}, 023525 (2005)
arXiv:astro-ph/0408081;

C.~Czaki, N.~Kaloper and J.~Terning,
Annals\ Phys.\ {\bf 317}, 410 (2005)
[arXiv:astro-ph/0409596];

S.~Tsujikawa and M.~Sami,
Phys.\ Lett.\ B {\bf 603}, 113 (2004)
[arXiv:hep-th/0409212];

P.~Gonzalez-Diaz and C.~Siguenza,
Nucl.\ Phys.\ {\bf B697}, 363 (2004)
[arXiv:astro-ph/0407421];

L.~P.~Chimento and R.~Lazkoz,
Phys.\ Rev.\ Lett.\ {\bf 91}, 211301 (2003)
[arXiv:gr-qc/0307111];
Mod.\ Phys.\ Lett.\ {\bf A19}, 2479 (2004)
[arXiv:gr-qc/0405020];

J.~Hao and X.~Li,
Phys.\ Lett.\ B {\bf 606}, 7 (2005)
[arXiv:astro-ph/0404154];


S.~Nesseris and L.~Perivolaropoulos,
Phys.\ Rev.\ D {\bf 70}, 123529 (2004)
[arXiv:astro-ph/0410309];

Z.~Guo,Y.~Piao, X.~Zhang, and Y.~Zhang,
Phys.\ Lett.\ B {\bf 608} 177 (2005)
[arXiv:astro-ph/0410654];

E.~Elizalde, S.~Nojiri, and S.~D.~Odintsov,
Phys.\ Rev.\ D {\bf 70}, 043539 (2004)
[arXiv:hep-th/0405034];

E.~Babichev, V.~Dokuchaev, and Yu.~Eroshenko,
Class.\ Quant.\ Grav. {\bf 22}, 143 (2005)
[arXiv:astro-ph/0407190];


X.~Zhang, H.~Li, Y.~Piao, and X.~Zhang,
arXiv:astro-ph/ 0501652;

M.~Bouhmadi-Lopez, and J.~Jimenez-Madrid,
JCAP {\bf 0505}, 005 (2005)
[arXiv:astro-ph/0404540];

Y.~Wei,
Mod.\ Phys.\ Lett.\ A {\bf 20}, 1147 (2005)
[arXiv:gr-qc/0410050];
arXiv:gr-qc/0502077;


V.~K.~Onemli and R.~Woodard,
Phys.\ Rev.\ D {\bf 70}, 107301 (2004)
[arXiv:gr-qc/0406098];

M.~Dabrowski and T.~Stachowiak,
arXiv:hep-th/0411199;

M.~C.~B.~Abdalla, S.~D.~Odintsov, and S.~Nojiri,
Class.\ Quant.\ Grav.\ {\bf 22}, L35 (2005)
[arXiv:hep-th/0409177];


R.-G.~Cai,
JCAP {\bf 0503}, 002 (2005)
[arXiv:hep-th/0411025];

R.-G.~Cai, H.-S.~Zhang, A.~Wang,
arXiv:hep-th/0505186;


I.~Ya.~Arefeva, A.~S.~Koshelev, and S.~Yu.~Vernov,
arXiv:astro-ph/0412619;

H.~Wei, R.-G.~Cai, and D.~Zeng,
arXiv:hep-th/0501160;

R.~Curbelo, T.~Gonzalez, and I.~Quiros,
arXiv:astro-ph/0502141;

B.~Gumjudpai, T.~Naskar, M.~Sami, and S.~Tsujikawa,
JCAP\ {\bf 0506}, 007 (2005)
[arXiv:hep-th/0502191];

H.~Lu, Z.~Huang and W.~Fang,
arXiv:hep-th/0504038.

S.~Nojiri, S.~D.~Odintsov, and M.~Sasaki,
Phys.\ Rev.\ D {\bf 71}, 123509 (2005)
[arXiv:hep-th/0504052];

M.~Sami, A.~Toropensky, P.~Trejakov, and S.~Tsujikawa,
arXiv:hep-th/0504154;

F.~Bauer,
arXiv:gr-qc/0501078;

J.~Sola and H.~Stefancic,
arXiv:astro-ph/0505133;

G.~Calcagni, S.~Tsujikawa, and M.~Sami,
arXiv:hep-th/0505193;

A.~Andrianov, F.~Cannata, and A.~Kamenshchik,
arXiv:gr-qc/0505087.

\bibitem{tsujikawa}
S.~Nojiri, S.~D.~Odintsov, and S.~Tsujikawa,
Phys.\ Rev.\ D {\bf 71}, 063004 (2005)
[arXiv:hep-th/0501025].

\bibitem{unification}
S.~Nojiri and S.~D.~Odintsov,
Phys.\ Lett.\ B {\bf 562}, 147 (2003)
[arXiv:hep-th/0303117];
Phys.Lett. B {\bf 571}, 1 (2003)
[arXiv:hep-th/0306212].

\bibitem{modified}
S.~Nojiri and S.~D.~Odintsov,
Phys.\ Rev.\ D {\bf 68}, 123512 (2003)
[arXiv:hep-th/0307288].

\bibitem{salvatore}
V.~F.~Cardone, A.~Troisi, and S.~Capozzielo,
arXiv:astro-ph/0506371.

\bibitem{EOS}
S.~Nojiri and S.~D.~Odintsov,
Phys.\ Rev.\ D {\bf 70}, 103522 (2004)
[arXiv:hep-th/0408170];

H.~Stefancic,
Phys.\ Rev.\ D {\bf 71}, 084024 (2005)
[arXiv:astro-ph/0411630];
arXiv:astro-ph/0504518;

M.~Szydlowski, W.~Godlowski, and R.~Wojtak,
arXiv:astro-ph/0505202.

\bibitem{inh}
S.~Nojiri and S.~D.~Odintsov,
arXiv:hep-th/0505215.

\bibitem{inflation}
Y.-S.~Piao, E.~Zhou,
Phys.\ Rev.\ D {\bf 68}, 083515 (2003)
[arXiv:hep-th/0308080];

P.~Gonzalez-Diaz and J.~Jimenez-Madrid,
Phys.\ Lett.\ B {\bf 596}, 16 (2004);

Y.~S.~Piao and Y.~Z.~Zhang,
Phys.\ Rev.\ D {\bf 70}, 063513 (2004);

A.~Anisimov, E.~Babichev, and A.~Vikman,
JCAP {\bf 0506}, 006 (2005)
arXiv:astro-ph/0504560.



\bibitem{CNO}
S.~Capozziello, S.~Nojiri and S.~D.~Odintsov,
arXiv:hep-th/0507182.



\bibitem{naftulin}
E.~Elizalde, S.~Naftulin, and S.~D.~Odintsov,
Phys.\ Rev.\ D {\bf 49}, 2852 (1994).

\bibitem{Li}
M.~Li,
Phys.\ Lett.\ B {\bf 603}, 1 (2004)
[arXiv:hep-th/0403127].

\bibitem{FEH}
Q-C.~Huang and Y.~Gong, JCAP {\bf 0408}, 006 (2004)
[arXiv:astro-ph/0403590];

B.~Wang, E.~Abdalla, and Ru-Keng~Su,
Phys.\ Lett.\ B {\bf 611}, 21 (2005)
[arXiv:hep-th/0404057];

Y.~S.~Myung,
arXiv:hep-th/0501023;
arXiv:hep-th/0502128;

K.~Enqvist and M.~S.~Sloth,
Phys.\ Rev.\ Lett. {\bf 93}, 221302 (2004)
[arXiv:hep-th/0406019];

S.~Hsu and A.~Zee,
arXiv:hep-th/0406142;

M.~Ito,
arXiv:hep-th/0405281;

P.~F.~Gonzalez-Diaz,
arXiv:hep-th/0411070;

Q.-G.~Huang, M.~Li,
JCAP {\bf 0503}, 001 (2005)
[arXiv:hep-th/0410095];


S.~Nobbenhuis,
arXiv:gr-qc/0411093;

Y.~Gong, B.~Wang, and Y.~Zhang,
arXiv:hep-th 0412218;

A.~J.~M.~Medved,
arXiv:hep-th/0501100;

X.~Zhang,
arXiv:astro-ph/0504586;

Yu.~Gong and Y.~Zhang,
arXiv:hep-th/0505175;

T.~Padmanabhan,
arXiv:gr-qc/0503107;

B.~Guberina, R.~Horvat, and H.~Stefancic,
JCAP {\bf 0505}, 001 (2005)
[arXiv:astro-ph/0503495];

B.~McInnes,
Nucl.\ Phys.\ B {\bf 718}, 55 (2005)
[arXiv:hep-th/0502209];

X.~Zhang and F.~Wu,
arXiv:astro-ph/0506310.

\bibitem{eli}
E.~Elizalde, S.~Nojiri, S.~D.~Odintsov, and P.~Wang,
Phys.\ Rev.\ D {\bf 71}, 103504 (2005)
[arXiv:hep-th/0502082].

\bibitem{amendola}
L.~Amendola,
Phys.\ Rev.\ D {\bf 62}, 043511 (2000);

W.~Zimdahl, D.~Pav\' on, and L.~P.~Chimento,
Phys.\ Lett.\ {\bf B521}, 133 (2001).



\bibitem{PZ}
D.~Pav\' on, W.~Zimdahl,
arXiv:gr-qc/0505020.

\bibitem{WGA}
B.~Wang, Y.~Gong and, E.~Abdalla,
arXive:hep-th/0506069.


\bibitem{verlinde}
E.~Verlinde,
arXiv:hep-th/0008140.

\bibitem{KeLi}
K.~Ke and M.~Li,
Phys.\ Lett.\ B {\bf 606}, 173 (2005)
[arXiv:hep-th/0407056];

M.~R.~Setare,
arXiv:hep-th/0405010.

\bibitem{BNOV}
I.~Brevik, S.~Nojiri, S.~D.~Odintsov, and L.~Vanzo,
Phys.\ Rev.\ D {\bf 70}, 043520 (2004)
[arXive:hep-th/0401073].

\end{thebibliography}
\end{document}